# Utility of Transient Testing to Characterize Thermal Interface Materials

B. Smith, T. Brunschwiler, and B. Michel
IBM Research GmbH, Zurich Research Laboratory
Säumerstrasse 4
8803 Rüschlikon, Switzerland

***Abstract*- This paper analyzes a transient method for the characterization of low-resistance thermal interfaces of microelectronic packages. The transient method can yield additional information about the package not available with traditional static methods at the cost of greater numerical complexity, hardware requirements, and sensitivity to noise. While the method is established for package-level thermal analysis of mounted and assembled parts, its ability to measure the relatively minor thermal impedance of thin thermal interface material (TIM) layers has not yet been fully studied. We combine the transient thermal test with displacement measurements of the bond line thickness to fully characterize the interface.**

## I. INTRODUCTION

Heat diffusion across the interface of two components in an electronic package is a well-studied topic in the thermal packaging field [1]. It is fundamental to any package, from conduction-only designs for handheld electronics [2, 3] to large datacenters featuring a complex combination of liquid and air fan cooling [4, 5]. The attention to thermal interfaces is justified as they account for up to 50% of the total thermal budget in some packages and directly influence product lifetime, performance, reliability, and power consumption.

Advanced thermal packages often use metal interfaces [6, 7], phase-change materials [8], or thermal interface materials made from highly particle-filled materials including greases and adhesives [9]. Despite this diversity, all thermal interface materials (TIM) are defined by their thermal resistance, $R_{TH}$ ($K \cdot cm^2/W$), and the optimal TIM for an application is a factor of thermal, mechanical, assembly, and cost considerations.

Acoustic mismatch, or the inability to couple the energy-carrying phonons of one surface to the interface material is refered to as contact resistance and is one contributor to the overall $R_{TH}$ of the TIM. More significant for interfaces with bondline thickness (BLT) greater than 1 μm is the resistance to heat diffusion within the TIM, which is captured simply by the Fourier equation for heat conduction:

$$Rth = \frac{BLT}{k_{eff}} \qquad (1)$$

where $k_{eff}$ is the effective thermal conductivity of the TIM. Thinner bondlines enable lower TIM resistance, limited by the uniformity of the surface and the TIM. As BLT decreases below 1μm, contact resistance (acoustic mismatch) can begin to dominate the overall $R_{TH}$ but for the systems considered in this work we will assume that this is negligible.

Most TIM characterization techniques induce 1-D heat flow normal to the interface and measure the steady thermal gradient through the system to extract the TIM's $R_{TH}$ [10]. The most popular setup involves four or six temperature sensors placed in-line with the heat flow, half on one side of the interface and half on the other. The discontinuity of the temperature gradient across the interface is attributed to the TIM, and Eq. 1 is used to extract effective conductivity if the BLT is known.

Another design, shown in Fig. 1, creates a direct analog to an electronic package by fabricating the heating element and temperature sensors on a chip and locating the TIM of interest directly in the thermal path between chip and heat sink [11]. In this setup, only the overall thermal resistance of the system can be measured directly so $R_{TH}$ of the TIM is extracted by either (1) characterizing the other resistances in the system separately and subtracting them from the measurement to obtain $R_{TH,TIM}$ or (2) measuring overall $R_{TH}$ for different BLT of the same TIM and associating the derivative of Eq. 1 to the change in overall system resistance. The main drawback of the first procedure is uncertainty in the thermal resistance of the liquid cooler and other parasitic components, whereas the second technique relies on very accurate sensing and control of BLT. In addition, any localized hot spots in the heater can decrease the accuracy in the interface characterization.

Alternately, transient or harmonic methods can isolate the influence of the TIM from the other components in the thermal path [12-14]. This can eliminate the inaccuracy in TIM properties due to cooler uncertainty, for example. Harmonic and pulse-based transient methods such as 3-omega and thermoreflectance have become the standards for thin film thermometry; however, the numerical fitting models and laser heat pulsing required limit their application in a product-like test fixture (Fig. 1), which is preferred for industrial package characterization.

One transient technique that is applicable to this test fixture is RC analysis using the thermal structure function [15, 16]. The technique calculates the RC spectrum of the high temporal resolution thermal response to a step function in power. Numerical deconvolution and transformation of the temporal response to the structure function can identify the magnitude of thermal conductivity and heat capacity of each region in the thermal path, thereby isolating the TIM from the rest of the system. The method may combine the advantages of transient testing with the flexibility of steady-state techniques.

The following sections describe the test fixture, review structure function techniques to determine TIM properties, and compare steady and transient results for a simple TIM system. Strengths and drawbacks of the transient technique are discussed including the ability of the structure function to resolve small $R_{TH}$ TIMs and the relevance of the technique to thermal packages with more complex geometry.

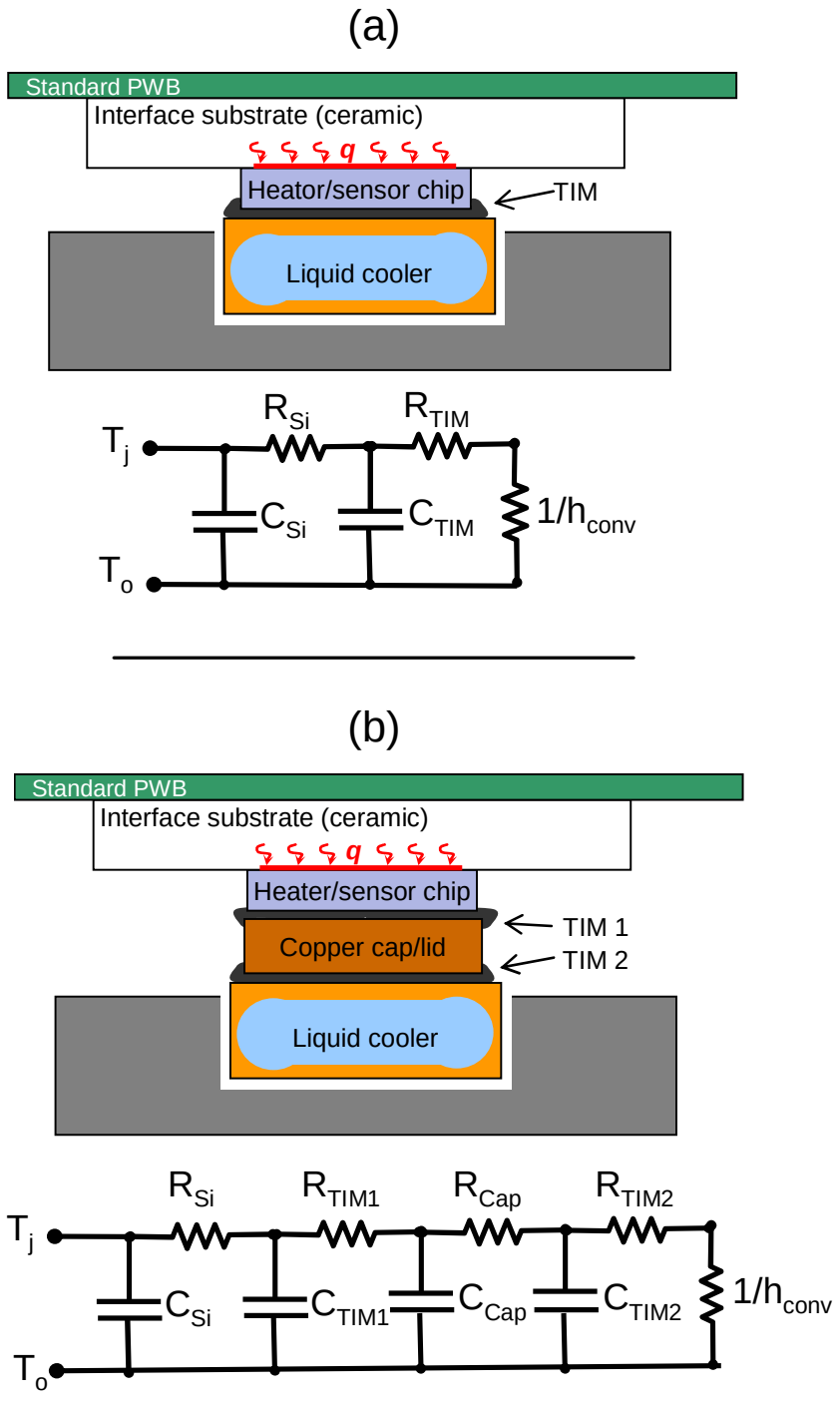

Fig. 1. Schematic layout of the test fixture with corresponding RC network model describing 1-D heat flow through the fixture; (a) refers to a simple heater-cooler arrangement with only one thermal interface, (b) includes a cap or lid as is often used as a heat spreader between cooler and heat source.

## II. EXPERIMENTAL SETUP

Fig. 1 shows the experimental fixture schematically. A pneumatic cylinder presses the IBM-designed thermal chip and chip carrier (ceramic substrate and PWB) module onto a liquid cooler (Micros, Inc.) fixed at 25°C. Serpentine resistors in the chip apply uniform heat with +/-5% uniformity across the entire surface. Nine RTD temperature sensors are distributed at the corners, center quadrants, and center of the chip to verify temperature uniformity. Induction-based displacement sensors (Mahr Inc., not shown) are mounted in the cooler housing and contact the ceramic substrate at the diagonals of the chip to measure BLT. The cooler and housing are mounted on a force sensor (ME-Meßsysteme GmbH.) to monitor the pressure/force applied during the TIM squeeze process. The final BLT after squeezing is a strong function of applied pressure and we use this information in the current study to control BLT.

Fig. 1 also shows two possible embodiments of the product-like fixture. Fig. 1a is analogous to a heat sink being directly attached to a chip and is the most straightforward design. Fig. 1b inserts a heat spreader between chip and cooler. This is a popular scheme to spread the heat laterally before interfacing to the cooler and the spreader is often fabricated as a cap sealing and protecting the chip in addition to spreading the heat. The design creates a second thermal interface ("TIM2") to consider in the thermal analysis. However, this paper focuses on the simple chip-TIM-cooler design of Fig. 1a. We address the applicability of the structure function-based transient test method to a variety of thermal package designs in later sections.

## III. EXPERIMENTAL RESULTS

### *A. Steady-state*

Steady-state measurements suggest TIM $k_{eff}$ = 4.8 W/mK +/- 10% using the method of varying BLT. A commercially-available thermal interface compound from Shin-Etsu Group was used as TIM for its high viscosity (easily-controllable BLT) and relevance to the industry. The results are consistent with previous measurements and other literature and confirm its exceptional thermal performance.

The thermal resistance of the 700 μm silicon chip is 0.05 $Kcm^2/W$ ($k_{Si}$= 141 W/mK). Therefore, the thermal resistance of the liquid cooler is 0.21 $Kcm^2/W$, or in terms of effective heat transfer coefficient, h = 48 $kW/m^2K$ – an excellent heat sink. Parasitic heat paths through the ceramic and the surrounding air were measured to be nearly two orders of magnitude higher in thermal resistance than the Si-TIM-cooler path so the assumption of 1-D heat flow captured in the RC network of Fig. 1a is valid. In practice, the parasitic resistances influence the cooler $R_{TH}$ prediction, which is not a critical data point in this study.

TABLE I
RELATIVE CONTRIBUTIONS TO TOTAL $R_{TH}$

| BLT | $R_{THsystem}$ | % Si | % TIM | % Cooler |
|---|---|---|---|---|
| 54.6 μm | 0.372 $Kcm^2/W$ | 13.3 | 30.5 | 56.2 |
| 15.2 μm | 0.288 $Kcm^2/W$ | 17.2 | 11.0 | 71.8 |
| 6.8 μm | 0.271 $Kcm^2/W$ | 18.3 | 5.2 | 76.5 |

Table 1 shows the relative contribution of each component in the heat path for the three BLTs measured. The relative magnitude of the TIM is an important metric to characterize the test's ability to resolve TIM properties. It shows that the cooler accounts for most of the resistance in the thermal path, and the TIM contribution drops to from 30.5 to 5.2% as the bond line is thinned. The thinnest bond line is at the limit of the accuracy of the displacement measurement. If a poorly-performing cooler were used, it is likely that the TIM contribution to over all $R_{TH}$ would be too small to accurately extract $k_{eff}$. This figure of merit will be formalized with respect to the transient test in later sections.

Finally, it is important to note that the steady state condition was reached within 2% approximately 20 seconds after the power step was applied. This is significant for determining the time window used for the transient measurements.

### *B. Transient*

Fig. 2 shows the temporal response to power step, displayed in log-time since the structure function relies on high temporal resolution (1 μs) at the beginning of the response. In general, the final (maximum) $R_{TH}$ predicted by the transient technique follows the steady predictions. Fig. 2 illustrates the noise associated with the power switching that is often encountered in the first 10 μs of the measurement. The normalized plots follow very similar profiles and the 4.2 μm case evolves slightly faster due to less thermal mass.

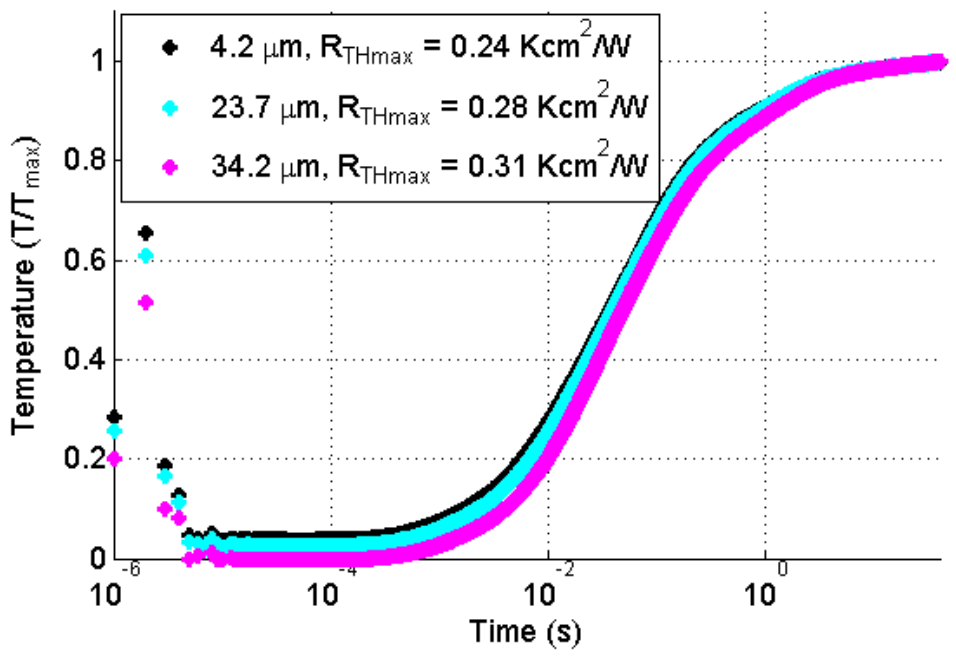

Fig. 2. Temporal response to power step for each BLT.

The temporal response in itself does not reveal TIM $R_{TH}$, however. It must be transformed into the structure function (SF), shown in Fig. 3a. The SF theory and mechanics as well as details regarding the hardware, switching noise, early transients, and numerical parameters are discussed elsewhere [16-18]. In brief, it is a kind of thermal history in the $R_{TH}$ domain, plotting the cumulative thermal resistance from source ($R_{TH}$ =0) to sink ($R_{TH}$=$R_{TH,steady}$) as observed by the temperature sensor which should be co-located with the heater for this analysis. The points in the $R_{TH}$ path where the heat capacitance changes suggest a change in material properties; i.e. an interface/junction between materials. Therefore, the components in the thermal path are distinguished in the $R_{TH}$ domain by observing the change in the SF.

Since we are primarily concerned with change in SF along the thermal path, the derivative of the structure function (DSF) with respect to $R_{TH}$ is computed numerically in Fig. 3b. Local maxima (“peaks”) in DSF denote material boundaries and the distance on the x-axis between boundaries is the thermal resistance of that material in the system.

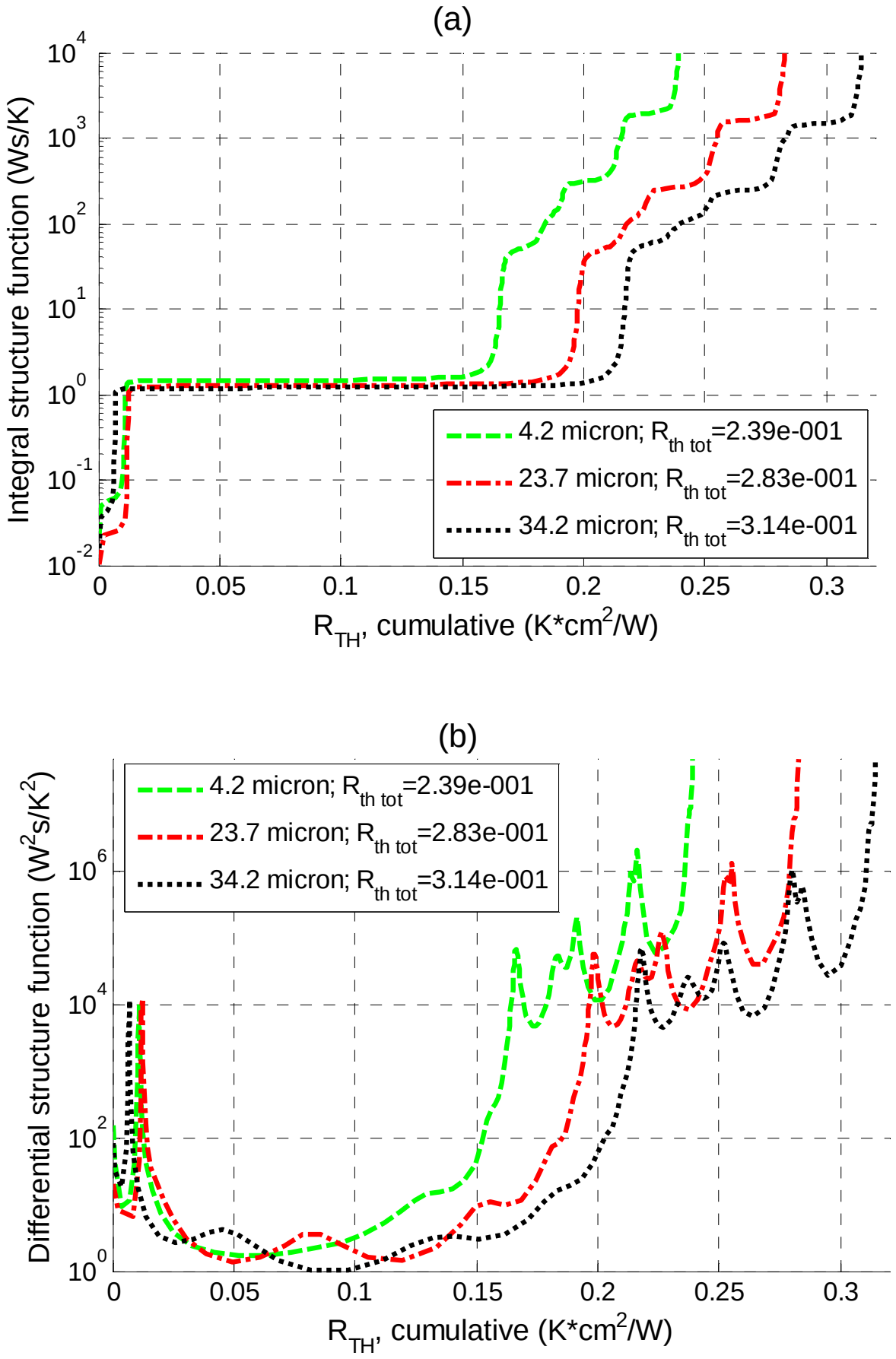

Fig. 3. Integral structure function (a) and differential structure function (b) for each BLT.

The main difficulty in analyzing SF or DSF is identifying and assigning the peaks in DSF to specific material boundaries. The most direct method is to follow the DSF path and assign boundaries corresponding to the thermal path. However, “false peaks” arise from early transients and the numerical transformations, and peaks toward the end of the resistance path tend to get “blurred out”. Proper test design can isolate 1-3 peaks in the part of the DSF that is not influenced by either early transients or blurring, however, the peak analysis is still highly subjective. Ongoing work involves comparing the experimental DSF with one obtained through analytical modeling of heat diffusion to identify and assign peaks more objectively.

Alternately, the variable BLT technique used for steady-state $k_{eff}$ measurement can be adapted for peak identification and quantification. Considering a magnification of the DSF from Fig. 3b in Fig. 4, it is clear that the highlighted peak moves from one BLT measurement to the next; therefore, it is the likely location of the TIM-cooler interface. Similar to the steady-state technique, the $k_{eff}$ is determined from measurement of $R_{TH}$ at these peaks, the BLT, and the derivative of Eq. 1. This technique yields $k_{eff,TIM}$ = 4.2 W/mK, +/- 15%, which is comparable to the steady results.

Conveniently, any set of matching peaks could be chosen to yield nearly identical $k_{eff,}$ since heat diffusion through the system is 1-dimensional. This would not be the case for a more complex geometry or some multi-TIM systems. In general, the final $R_{TH}$ predicted by the transient measurement scales with the steady state measurements well, considering different BLTs were used for each. Like the static results, however, the error/variation of the $k_{eff}$ prediction with the rest of the data set was worse as the BLT was reduced to the uncertainty of the displacement measurement. The error of the prediction with respect to BLT and magnitude of TIM $R_{TH}$ is discussed in a later section.

Single transient measurements are sensitive to noise both in the high frequency regime (1 MHz to 10 kHz and in the low frequency regime (1 Hz to 0.01 Hz). Systematic errors can be induced by the power measurement and by the inability of the power supply to provide and exact rectangular power step in 1 µs. Other problems are the slow change of resistance of the heater that starts to change its temperature as a consequence of the transient experiment. The averaging procedure to combine the results of multiple datasets improves the accuracy of the TIM peaks in the differential structure function graph by eliminating both high frequency noise and low-frequency changes that are not correlated with the power step. In this study, each plotted transient results is actually the average over five power step cycles and the variation among results was generally less than 8%.

Two factors that strongly influence the shape and peaks of the DSF are the numerical parameters of the deconvolution and Foster-Cauer network transformations and the time window of the measurement. A detailed discussion of the role of the numerical parameters is outside the scope of this work; however, we note that the Bayesian iteration number can be used to "sharpen" the peaks for better quantitative identification at the cost of false peaks appearing in the DSF, as shown in Fig. 4. A moderate iteration number of 4096 creates sharp peaks at the TIM-cooler interface without introducing false peaks.

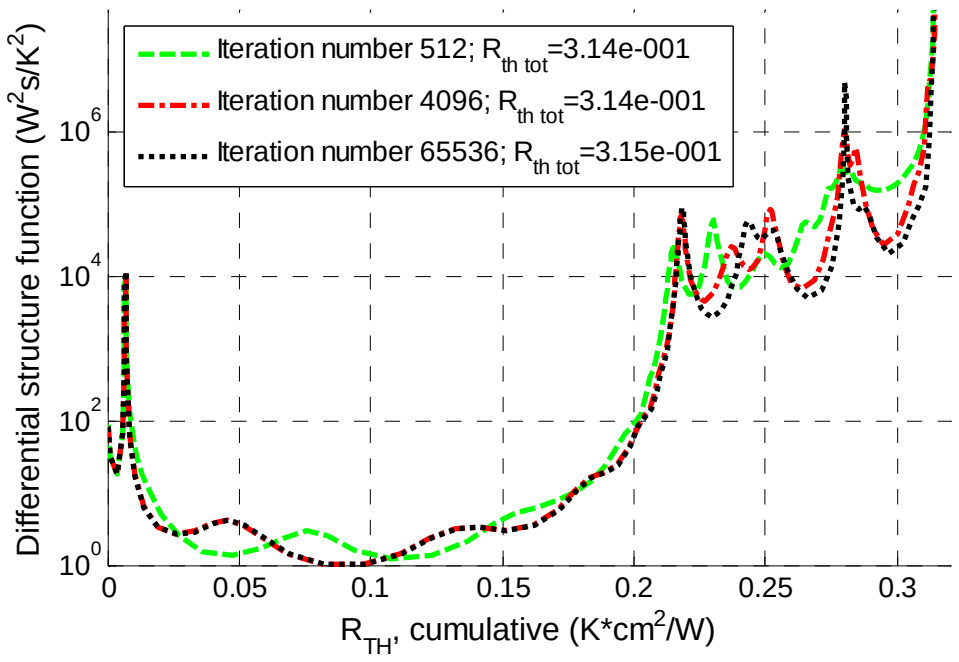

Fig. 4. Effect of Bayesian iteration number on DSF.

The influence of the measurement time is shown in Fig. 5 for the case of BLT= 34 µm. Observing from the steady tests that thermal steady state occurred in approximately 20 seconds, the transient measurement was run at multiples of ½, 1, 2 and 4 of this time. The total measurement time does not affect the short-time resolution of the measurement; however it does affect the resolution of the structure function in the $R_{TH}$ domain. At long measurement times, the early/middle transients where TIM lies are lost as the structure function must span a larger total $R_{TH}$ range. Too short a time window can cause incorrect scaling of the $R_{TH}$ and therefore incorrect quantitative results. Here, the prediction of overall $R_{TH}$ was quite consistent for all the measurements but the peaks converged best for the long time, 2 and 4 times the approximated time constant (see inset, Fig. 5). Practically speaking, a shorter measurement is ideal but 2 times the observed time constant is reasonable.

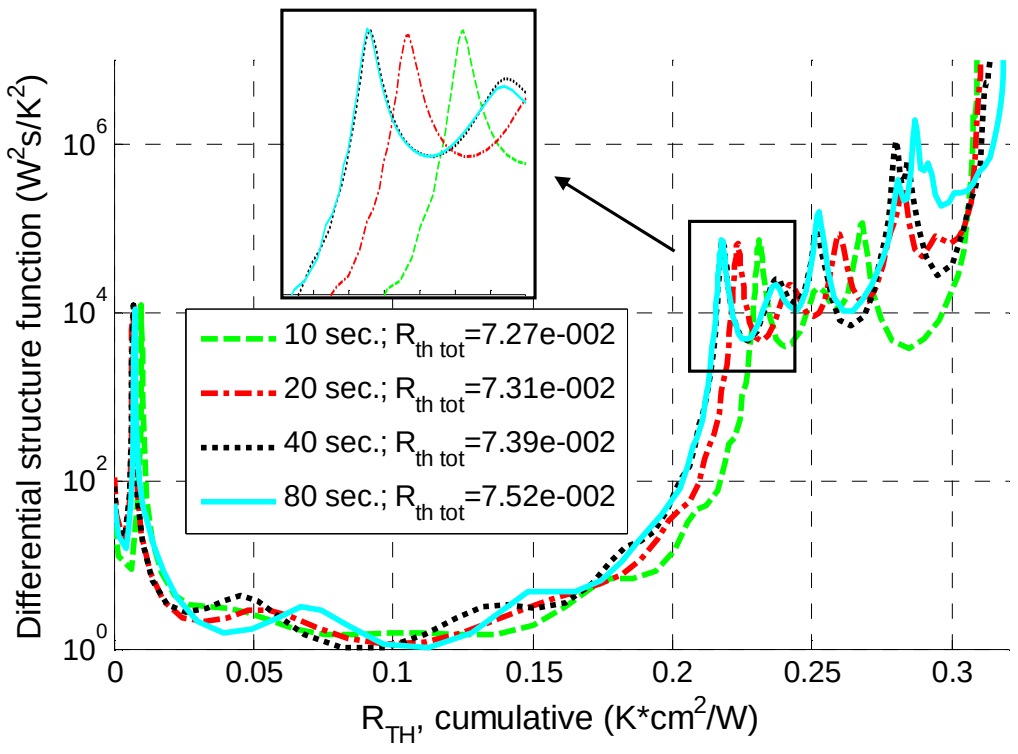

Fig. 5. Influence of measurement time on DSF.

## IV. General Figures of Merit

Two figures of merit are developed to validate the technique using both the experimental results and analytical models that will be formally presented in later studies.

First, the relative thermal resistance of the TIM compared to the overall resistance of the package determines the ability of the numerical deconvolution and structure function calculations to resolve the TIM peaks accurately. When the TIM comprises more than 70% of the total resistance, the peak is pushed close to the end of the thermal path and smeared out; however, good agreement with the expected peaks occurs when the TIM is greater than 25% of the total thermal resistance. Fig. 6 expresses this concept by examining the error of the location of the peak ($R_{TH}$ location) with respect to a known analytical model input as the TIM BLT is varied to sweep the scale from 10% of the total resistance to 90%. Ideally, the test setup should ensure that the TIM is the dominant thermal resistance, however, this is increasingly difficult for thin, high performance TIMs. As the experimental case studied in this work confirms, even the best liquid coolers have significantly higher $R_{TH}$ than a good TIM at less than 10 μm BLT.

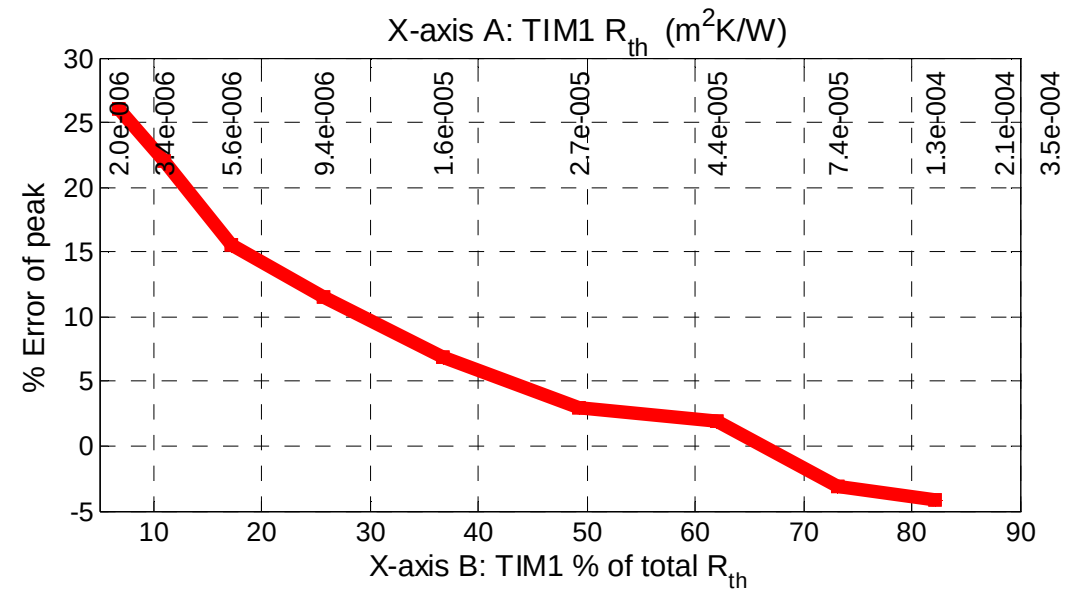

Fig. 6. Error from peak analysis with respect to the overall $R_{TH}$ contribution of the TIM.

Secondly, the location of the TIM interface in the overall thermal path from heater to cooler determines the extent to which the peak can be distinguished from the early transients in the response and from the “smearing” that occurs close to the end of the path (the cooler). Analytical modeling showed that the peak in DSF is sharp and distinct when the interface lies 20% of the total thermal path away from the heater but 30% from the cooler, yielding a window of maximum peak distinction. This drives the test design and may limit the accuracy of transient TIM characterization of more complex designs involving more than one TIM and/or a spreading element like that shown in Fig. 1b.

## V. Conclusions

The variable BLT technique applied to transient characterization through the structure function can provide a quantitative assessment of TIM $R_{TH}$ and $k_{eff}$. It is preferred over a direct peak-matching scheme; however, it may be only marginally more useful than static test methods when the hardware and software overhead of transient testing are considered. Ongoing work includes fitting analytical RC models in the time and structure function domains to the experimental data to characterize the TIM directly and without using the variable bond line method. This would speed up the time required to characterize a TIM and eliminate the need to measure BLT if $R_{TH,TIM}$ is the only required data point. Experimental validation of the figures of merit and test cases using other electronic package geometries are also underway. Work is also underway to improve the signal-to-noise ratio and to reduce systematic errors in transient measurements.

## Acknowledgment

The authors thank Hugo Rothuizen, Urs Kloter, Ryan Linderman, Evan Colgan, and Paul Seidler for discussions and support.